\documentclass[journal=jacsat,manuscript=article]{achemso}  
\usepackage{amsfonts}
\usepackage{amssymb}
\usepackage{graphicx}
\usepackage{amsmath}
\usepackage{tikz}
\usepackage{adjustbox}
\usepackage[normalem]{ulem}
\usetikzlibrary{shapes.geometric, arrows}

\usepackage[english]{babel}
\usepackage{color}
\usepackage[version=3]{mhchem} 
\usepackage{natbib}
\usepackage{url}
\usepackage[colorlinks,citecolor=black,urlcolor=blue,linkcolor=black]{hyperref} 
\usetikzlibrary{arrows,shapes,positioning,shadows,trees}

\newcommand{\olr}[1]{{\color{red}{}}}

\newcommand{\osvp}[1]{{\color{blue}{}}}

\author{Sebastian V. Pios}
\affiliation{Zhejiang Laboratory, Hangzhou 311100, China}
\author{Maxim F. Gelin}
\affiliation{School of Science, Hangzhou Dianzi University, Hangzhou 310018, China}
\author{Wolfgang Domcke}
\affiliation{Department of Chemistry, Technical University of Munich, D-85747 Garching, Germany}
\author{Lipeng Chen}
\affiliation{Zhejiang Laboratory, Hangzhou 311100, China}
\email{chenlp@zhejianglab.com}

\title{Simulation of Pump-Push Molecular Dynamics in the Heptazine-H\textsubscript{2}O Complex}

\begin{document}


\begin{abstract}
Pump-push-probe spectroscopy was employed for the exploration of charge-separation processes in organic photovoltaic blends as well as for proton-coupled electron-transfer (PCET) reactions in hydrogen-bonded complexes of tri-anisole-heptazine with substituted phenols in organic solvents. 
In the present work, the electron and proton transfer dynamics driven by a femtosecond pump pulse and a time-delayed femtosecond push pulse has been studied with \textit{ab initio} on-the-fly nonadiabatic trajectory calculations for the hydrogen-bonded heptazine-H\textsubscript{2}O complex. 
While the dynamics following the pump pulse is dominated by ultrafast radiationless energy relaxation to the long-lived lowest singlet excited state (S\textsubscript{1}) of the heptazine chromophore with only minor PCET reactivity, the re-excitation of the transient S\textsubscript{1} population by the push pulse results in a much higher PCET reaction probability. 
These results illustrate that pump-push excitation has the potential to unravel the individual electron and proton transfer processes of PCET reactions on femtosecond time scales.
\end{abstract}

%
Melon,\cite{liebig_1834} graphitic carbon nitride\cite{wang_angewandte_2012} and numerous related polymeric carbon nitrides\cite{lau2022tour,kumar2023multifunctional,pelicano2024metal} have received great attention in the renewable energy research community as highly promising photocatalysts for the evolution of molecular hydrogen and oxygen from water under UV/vis irradiation.\cite{wang2009metal,ong_chemrev_2016,WEN2017}
Despite this extensive research, the elementary reaction mechanisms involved in the photocatalytic hydrogen and oxygen evolution reactions, such as exciton dissociation, charge carrier migration and the oxidation of water, are still not well understood. 
The poorly constrained stoichiometries and atomistic structures of these polymeric materials are detrimental to the precise characterization of the primary electron and proton transfer processes involved in the photocatalytic hydrogen and oxygen evolutions reactions.

In an alternative approach, the photochemistry of heptazine (tri-s-triazine) - the molecular building block of most carbon nitrides - has been explored in aqueous solution and in organic solvents.\cite{domcke2020photooxidation,schlenker_overview2021}
While the heptazine (Hz) molecule is chemically unstable due to rapid hydrolysis in the presence of water,\cite{halpern_hz_1984} a derivative of Hz, trianisole-heptazine (TAHz), was found to be chemically and photochemically highly stable.\cite{rabe_tahz_2018} 
It was shown by Schlenker and coworkers that 365~nm LED irradiation of TAHz in aqueous solution results in the liberation of OH radicals generated by an excited-state proton-coupled electron-transfer (PCET) reaction.\cite{rabe_tahz_2018}
Adopting para-substituted phenols as model substrates in organic solvents, Schlenker and coworkers systematically characterized the PCET reactions in hydrogen-bonded TAHz-phenol complexes with kinetic measurements and accompanying \textit{ab initio} electronic-structure calculations of reaction paths and potential-energy barriers.\cite{rabe2019barrierless} 
These studies provided a generalizable picture of barrier-controlled excited-state PCET reactions in hydrogen-bonded complexes of a Hz derivative with protic substrate molecules.\cite{schlenker_overview2021} 

Additional insight into the nonadiabatic excited-state processes involving non-reactive locally-excited (LE) states and reactive charge-transfer (from phenol to TAHz) states was obtained by time-resolved pump-push-probe spectroscopy.\cite{corp_jpcc_2020} 
In this experiment, the TAHz-phenol complex was first excited to the lowest bright $^{1}\pi\pi^{*}$ state of TAHz by a UV (365~nm) pump pulse.
While part of the excited complexes underwent PCET (from phenol to TAHz) via reactive charge-transfer (CT) states, most of the excited electronic population relaxed to the long-lived S\textsubscript{1} ($\pi\pi^{*}$) state of the Hz core. 
After a time delay of 6~ps, an infrared (1150~nm) push pulse was applied to excite the S\textsubscript{1} population to an energy region in which nonadiabatic interactions of the LE states with reactive CT states are prevalent. 
The enhanced PCET reactivity was monitored with a supercontinuum probe pulse as a persistent dip of the $\Delta$OD (difference of optical density) signal.\cite{corp_jpcc_2020} 
Such pump-push-probe experiments were performed earlier for the exploration of the mechanisms of charge separation in organic polymers.\cite{tan2013suppressing, mangold2014control, zhang2017energetics, paterno2019pump}

The present work builds on previous computational studies of the photoreactivity of the hydrogen-bonded Hz$\cdots$H\textsubscript{2}O complex with \textit{ab initio} on-the-fly nonadiabatic trajectory surface-hopping calculations.\cite{xiang_heptazine2021,pios_hz_h2o_pumpprobe} 
In the excited-state PCET reaction, an electron and a proton are transferred from the water molecule to one of the nitrogen atoms of the Hz molecule along the hydrogen bond of the Hz$\cdots$H\textsubscript{2}O complex, resulting in HzH and OH radicals in their electronic ground states.\cite{ehrmaier_hz_2017}
These quasi-classical dynamics simulations provided insight into the interplay of nonadiabatic transitions between LE and CT states, electron transfer, proton transfer and vibrational energy relaxation.\cite{xiang_heptazine2021,pios_hz_h2o_pumpprobe}  
Within the explored time window of 100~fs, a few percent of the trajectories encountered a conical intersection of the S\textsubscript{1} state with the electronic ground state, while the majority of trajectories relaxed via intermediate $^{1}n\pi^{*}$ states to the long-lived dark S\textsubscript{1} ($\pi\pi^{*}$) state of Hz. 
Although a significant part of the vibrationally hot population of the S\textsubscript{1} state likely overcomes the barrier for H-atom transfer to Hz or tunnels through this barrier (height estimated as $\approx$ 1.0~eV \cite{ehrmaier_jpca_2020_molecular}), the time scales of these processes are too long to be simulated with \textit{ab initio} on-the-fly trajectory calculations.

Inspired by the experimental work of Schlenker and coworkers on TAHz-phenol complexes,\cite{schlenker_overview2021} we explored \textit{in silico} the effect of a push pulse on the photochemistry of the Hz$\cdots$H\textsubscript{2}O complex. 
As in Refs. \cite{xiang_heptazine2021,pios_hz_h2o_pumpprobe}, the pump pulse is assumed to excite the Hz$\cdots$H\textsubscript{2}O complex within an energy window of 0.1 eV at the maximum of the absorption spectrum of the bright S\textsubscript{5}/S\textsubscript{6} excited state (4.29~eV). 
The thus generated trajectories are propagated up to 100~fs. 
At this time, the radiationless relaxation of the higher excited singlet states to the long-lived S\textsubscript{1} ($\pi\pi^{*}$) state of Hz is essentially finished. 
The resulting hot vibrational distribution in the S\textsubscript{1} state provides the probability distribution of positions and momenta from which the initial conditions simulating the excitation by the push pulse are sampled according to their oscillator strengths within a specified energy window. 
The ensemble of trajectories excited by the push pulse is propagated for another 100~fs (up to 200~fs in total). 
The details of the nonadiabatic quasi-classical trajectory simulations are described in the Supporting Information (SI).

\begin{figure}
    \centering
    \includegraphics[width=0.6\linewidth]{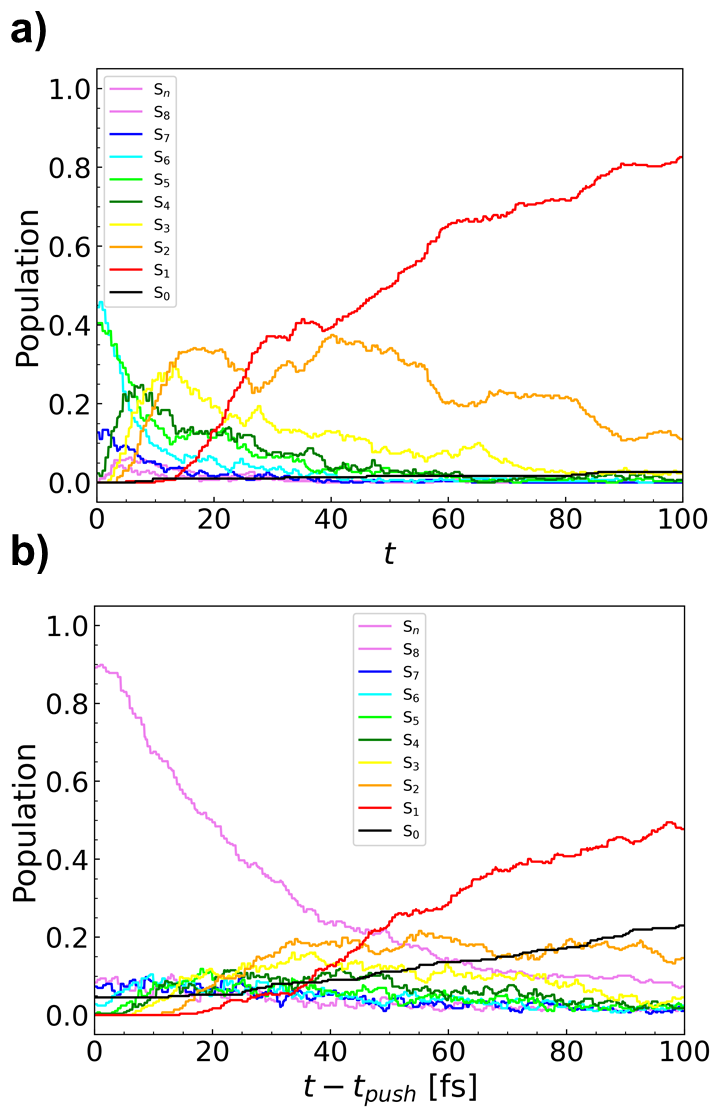}
    \caption{(a) Population probabilities of the adiabatic singlet states in the Hz$\cdots$H\textsubscript{2}O complex as a function of \textit{t} after the pump pulse. (b) Population probabilities of the adiabatic singlet states as a function of  \textit{t} - \textit{t}\textsubscript{push} after reexcitation by the push pulse (see legend for the color code). The populations of higher-lying electronic states are displayed collectively and labeled as S\textsubscript{\textit{n}}. The data in (a) are taken from Ref.~\cite{pios_hz_h2o_pumpprobe}.}
    \label{fig:adia_pop}
\end{figure}

%
The pump pulse centered at 4.29~eV with a width of 0.1~eV populates at \textit{t} = 0~fs primarily the quasi-degenerate bright S\textsubscript{5}/S\textsubscript{6} electronic states of Hz (see Section S2.1 of the SI). 
The S\textsubscript{7} ($n\pi^{*}$) state also acquires some population due to vibronic mixing with the S\textsubscript{5}/S\textsubscript{6} states.
The time evolution of the populations of the adiabatic electronic states up to 100~fs is displayed in Figure~\ref{fig:adia_pop}a. 
The initial populations of the S\textsubscript{5}, S\textsubscript{6} and S\textsubscript{7} states decay on an ultrafast time scale of about 10~fs, while the population of the S\textsubscript{2} ($n\pi^{*}$) state rises on the same time scale. 
After a delay of about 15~fs, the population of the S\textsubscript{1} ($\pi\pi^{*}$) state rises quickly and subsequently more slowly. 
At \textit{t} = 100~fs, more than 80\% of the initially excited electronic population is in the S\textsubscript{1} state, see Figure~\ref{fig:adia_pop}a. 
Upon close inspection of Figure~\ref{fig:adia_pop}a one can see that about 5\% of the population are in the S\textsubscript{0} state (black curve) at \textit{t} = 100~fs.

%
\begin{figure}
    \centering
    \includegraphics[width=0.8\linewidth]{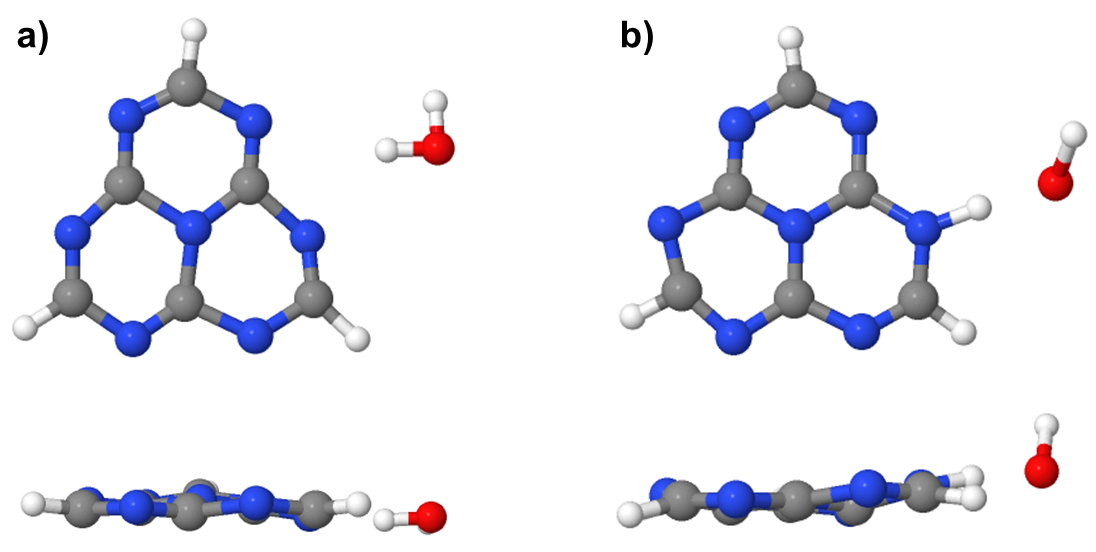}
    \caption{(a) Top and side view of the structure of a non-reactive conical intersection. (b) Top and side view of the structure of a reactive conical intersection.}
    \label{fig:rep_geos}
\end{figure}

The simulated push pulse at t\textsubscript{\textit{push}} = 100~fs is centered at an energy of 2.8~eV with a width of $\Delta E\approx$ 3.7~eV, corresponding to a duration of $\tau =$ 1~fs (see Section S2.3 of the SI). 
It excites the ensemble of trajectories in the S\textsubscript{1} state to higher excited singlet states S\textsubscript{n} with probabilities given by the S\textsubscript{1} - S\textsubscript{n} oscillator strengths. 
The resulting electronic populations and their time evolution are shown in Figure~\ref{fig:adia_pop}b. 
At \textit{t} = \textit{t}\textsubscript{\textit{push}}, more than 80\% of the electronic population are in higher excited adiabatic electronic states denoted collectively as “S\textsubscript{n}” in Figure~\ref{fig:adia_pop}b. 
Lower lying states, such as S\textsubscript{6}, S\textsubscript{7}, S\textsubscript{9}, are populated with a few percent. 
Figure~\ref{fig:adia_pop}b reveals that the “S\textsubscript{n}” population decays on a time scale of about 30~fs and that the S\textsubscript{1} state is re-populated on a similar time scale. 
Notably, the population of the S\textsubscript{0} state has grown to about 20\% at \textit{t} – \textit{t}\textsubscript{\textit{push}} = 100~fs. 
The analysis of the nuclear geometries of the trajectories terminating in the S\textsubscript{0} state reveals that 56\% of these trajectories are nonreactive (that is, no proton transfer from H\textsubscript{2}O to Hz takes place) and 44\% are reactive, representing HzH$\cdots$OH biradicals in their electronic ground states.
The excitation of the S\textsubscript{1} population by the push pulse thus increases the proton-transfer reaction probability by about an order of magnitude. 
The molecular geometry of a representative non-reactive product in the S\textsubscript{0} state is illustrated in Figure~\ref{fig:rep_geos}a, while the molecular geometry of a representative reactive product is shown in Figure~\ref{fig:rep_geos}b.
The non-reactive product geometry exhibits strong in-plane deformations with only minor out-of-plane displacements.
The reactive product geometry is similar to previously reported geometries\cite{pios_hz_h2o_pumpprobe} with the HzH radical exhibiting both, in-plane and out-of-plane, deformations.

While the time-dependent electronic populations provide information on the radiationless electronic decay dynamics after impulsive excitation of Hz by the pump pulse and by the push pulse, they do not inform on the dynamics of the mobile proton along the intermolecular hydrogen bond. 
To visualize the essential features of the excited-state PCET reaction, we constructed two-dimensional dynamic maps of the ensemble of trajectories. 
As proposed by Huang and Domcke,\cite{huang2022ab} the OH distance R\textsubscript{OH} (which serves as the proton-transfer coordinate) and the distance R\textsubscript{ON} of the donor and the acceptor of the proton is monitored for all trajectories propagating in the S\textsubscript{1} state. 
The electronic potential energy (relative to the local minimum of the S\textsubscript{1} potential-energy surface) is color-coded on a scale varying from dark blue for low potential energies to red for the highest potential energies. 
The trajectories are monitored with a time step of 0.5~fs. 
The encounter of a conical intersection of the S\textsubscript{1} potential-energy surface with the S\textsubscript{0} potential-energy surface (which terminates the trajectory) is marked by a cross at the position of the conical intersection in R\textsubscript{OH} - R\textsubscript{ON} space. 
Since the energy of each trajectory is conserved, low electronic potential energies indicate high nuclear kinetic energies and vice versa.

\begin{figure}
    \centering
    \includegraphics[width=0.55\linewidth]{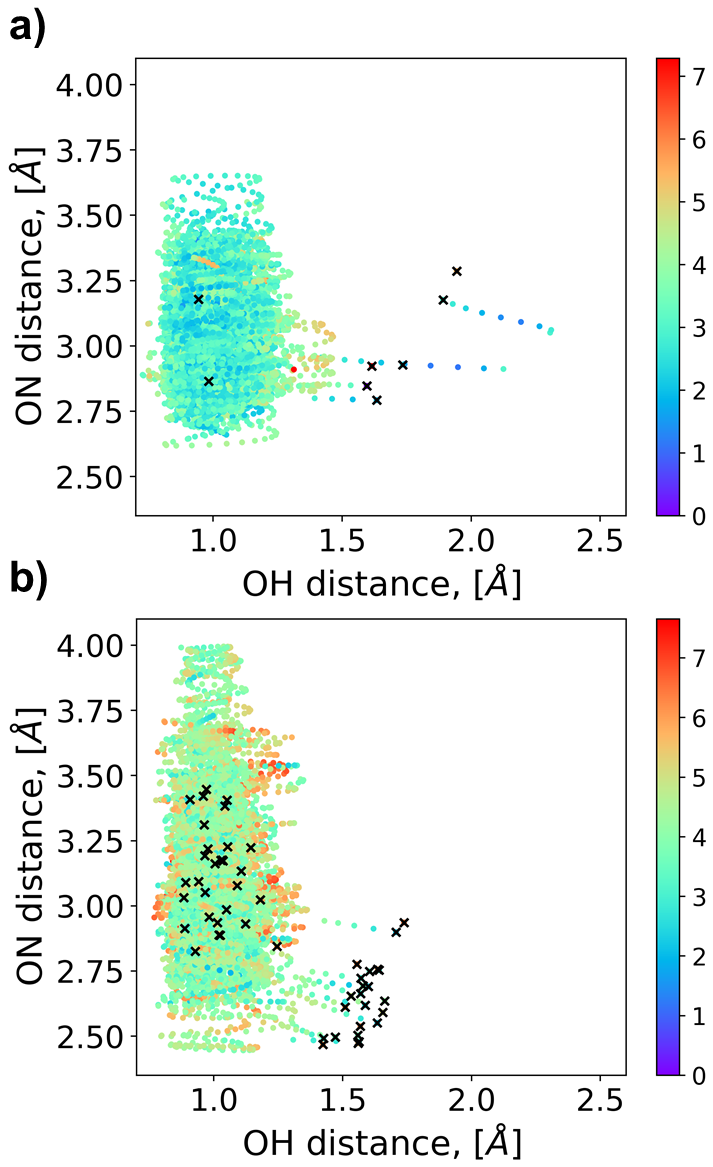}
    \caption{(a) Dynamic map (projection of the trajectories onto the plane of the OH and ON distances) covering the first 100~fs following excitation by the pump pulse.
     The data are taken from Ref.~ \cite{pios_hz_h2o_pumpprobe}. (b) Dynamic map covering the first 100~fs after re-excitation by the push pulse. The energy is given relative to the local  S\textsubscript{1} minimum. The crosses denote terminations of trajectories at S\textsubscript{1}-S\textsubscript{0} CIs.}
    \label{fig:dynamic_map1}
\end{figure}

Figure~\ref{fig:dynamic_map1}a displays the dynamic map of the trajectories initiated at \textit{t} = 0~fs by the pump pulse. 
The potential energy of the trajectories is in the range 2.0 – 4.0~eV. 
The OH distance is confined to the interval 0.9~\AA – 1.3~\AA\space and the ON distance varies from 2.7~\AA\space to 3.6~\AA. 
This part of the dynamic map represents radiationless electronic decay dynamics in the Hz chromophore from the bright S\textsubscript{5}/S\textsubscript{6} states to the S\textsubscript{1} state. 
The region R\textsubscript{OH} $>$ 1.5~\AA\space represents trajectories which have undergone proton transfer. 
Just a few trajectories enter the proton-transfer region. 
Their dynamics is driven by the (diabatic) CT state and they terminate at an S\textsubscript{1}-S\textsubscript{0} conical intersection. 
It is seen that the proton-transfer dynamics is strongly correlated with the donor-acceptor distance R\textsubscript{ON}, which is typical for proton-transfer reactions. 
The reduction of the ON distance lowers the barrier of the proton-transfer reaction. 
Figure~\ref{fig:dynamic_map1}a illustrates that the PCET reaction is a rare event in the Hz$\cdots$H\textsubscript{2}O complex due to the very efficient radiationless relaxation within the LE excited states of Hz and the rather high barrier for proton transfer ($\approx$ 1.0~eV) on the S\textsubscript{1} potential-energy surface. 
Closer inspection reveals that the two conical intersections encountered at R\textsubscript{OH} $<$ 1.0~\AA\space in Figure~\ref{fig:dynamic_map1}a result from the transfer of the second proton of H\textsubscript{2}O (which is not monitored in the dynamic map). 
Thus all encounters of S\textsubscript{1}-S\textsubscript{0} conical intersections in Figure~\ref{fig:dynamic_map1}a occur through a proton-transfer reaction and the branching ratio for proton transfer is rather low. 
It should be kept in mind, however, that only proton transfer on the S\textsubscript{1} surface is recorded. 
Proton transfer processes occurring in higher electronic states are not recorded in the dynamic map of Figure~\ref{fig:dynamic_map1}a.

The dynamic map shown in Figure~\ref{fig:dynamic_map1}b represents the dynamics after re-excitation of the Hz chromophore by the push pulse.
The overall change of color reflects the higher electronic potential energy available after the push pulse. 
The range of the majority of trajectories in the OH distance is about the same as in Figure~\ref{fig:dynamic_map1}a, while the range in the ON distance is increased, from 2.5~\AA\space to 4.0~\AA. 
It is obvious that the number of encounters of S\textsubscript{1}-S\textsubscript{0} conical intersections is strongly increased compared to Figure~\ref{fig:dynamic_map1}a. 
The electronic potential energy imparted by the push pulse is rapidly converted into nuclear kinetic energy. 
The resulting large-amplitude nuclear motion of the Hz chromophore facilitates access to S\textsubscript{1}-S\textsubscript{0} conical intersections. 
The conical intersections encountered at OH distances $<$ 1.3~\AA\space lead to direct internal conversion in the Hz chromophore. 
The conical intersections encountered at OH distances $>$ 1.3~\AA\space represent PCET reactions leading to HzH and OH radicals. 
The increased nuclear kinetic energy on the S\textsubscript{1} energy surface also increases the probability of overcoming the barrier for proton transfer (H-atom tunneling is not taken into account in classical trajectory dynamics). 
The proton-transfer yield is thus significantly enhanced by the push pulse at the cost of loss of electronic population from the reservoir of the S\textsubscript{1} state by competing direct internal conversion to the S\textsubscript{0} state.

It is eye-catching that the reactive conical intersections in Figure~\ref{fig:dynamic_map1}a as well as in Figure~\ref{fig:dynamic_map1}b appear to be approximately located along a line which is tilted in the OH-ON plane, in contrast to the nonreactive conical intersections in Figure~\ref{fig:dynamic_map1}b, which are located in a rectangular region with OH distances around 1.0~\AA\space and ON distances between 2.8~\AA\space and 3.5~\AA. 
The clear correlation of the OH and ON distances for the reactive conical intersections suggests that they may occur at a fixed NH distance of the Hz$\cdots$H\textsubscript{2}O complex. 
This expectation is confirmed by the dynamic maps constructed as functions of the NH and ON distances which are displayed in Figures~S3a and S3b. 
It is seen that the reactive conical intersections are aligned on a vertical line close to 1.0~\AA\space in the NH-ON plane. 
The nonreactive conical intersections, on the contrary, are broadly distributed in the NH distance from about 2.0~\AA\space to 4.0~\AA. 
The formation of the covalent NH bond of the HzH radical is thus critical for the occurrence of reactive S\textsubscript{1}-S\textsubscript{0} conical intersections in the Hz$\cdots$H\textsubscript{2}O complex.

%
%
Previous simulations of the photophysics and photochemistry of the Hz$\cdots$H\textsubscript{2}O complex have shown that excitation of the bright S\textsubscript{5}/S\textsubscript{6} states of the Hz chromophore results in a chain of ultrafast nonadiabatic transitions which involve intermediate LE $n\pi^{*}$ states and a CT state which can promote proton transfer from H\textsubscript{2}O to Hz.\cite{xiang_heptazine2021,pios_hz_h2o_pumpprobe} 
The branching ratio for proton transfer was found to be relatively low. 
The vast majority of electronic population is accumulated in the long-lived S\textsubscript{1} state of Hz. 
A short and suitably delayed push pulse can re-excite the transient population of the S\textsubscript{1} state to the energy region in which the bifurcation of reactive (proton transfer) and nonreactive (internal conversion in Hz) processes occur. 
Because the S\textsubscript{1} state is vibrationally hot due to the excess energy of the radiationless transitions within Hz, the re-excited system exhibits enhanced photochemical reactivity on ultrafast time scales which may be detected by an ensuing probe pulse as loss of the population of the S\textsubscript{1} state.

In the present work, we simulated the pump-induced dynamics as well as the pump-push-induced dynamics with \textit{ab initio} on-the-fly trajectory simulations to obtain further insight into the excited-state PCET reaction in the Hz$\cdots$H\textsubscript{2}O complex. 
The simulated pump pulse has been chosen to excite the Hz chromophore within a window of 0.1~eV at the maximum of the absorption spectrum of the bright S\textsubscript{5}/S\textsubscript{6} states. 
The dynamic map of the trajectories reveals that only a small fraction of trajectories ($\approx$ 5\%) encounter an S\textsubscript{1}-S\textsubscript{0} conical intersection via proton transfer. 
A short ($\tau$ = 1~fs) and energetically broad ($\Delta E\approx$ 3.7~eV) push pulse has been applied at t\textsubscript{0} = 100~fs and the dynamics was followed for another 100~fs. 
While the re-excited electronic states exhibit a somewhat slower ($\approx$ 50~fs) radiationless relaxation back to the S\textsubscript{1} state, the branching ratio for proton transfer is increased to about 20\% at the end of the simulation time (200~fs). 
The dynamic map of the trajectory dynamics following the push pulse also reveals enhanced S\textsubscript{1}-S\textsubscript{0} internal conversion without proton transfer. 
The dip in the transient S\textsubscript{1} population caused by the push pulse thus is due to enhanced proton-transfer reactivity as well as enhanced local internal conversion dynamics within the Hz chromophore.

Our simulations show that pump-push excitation is a suitable scenario for the unraveling of competing ultrafast photophysical and photochemical processes in hydrogen-bonded systems. 
Additional information could be retrieved by varying the duration and the delay time of the push pulse. The complete simulation of pump-push-probe spectroscopic signals is the subject of work in progress.

\section{Supporting Information}
Electronic structure calculations, initial conditions and push pulse simulation, quasi-classical nonadiabatic molecular dynamics, excited state absorption profile, distribution of vertical excitation energies, dynamic maps with respect to NH and ON distance
\begin{acknowledgement}
S.V.P. acknowledges support from the National Natural Science Foundation of China (No.~W2433024, No.~22473101). 
L.P.C. support from the National Natural Science Foundation of China (No.~ 22473101).
M.F.G. acknowledges support from the National Natural Science Foundation of China (No.~22373028). 
\end{acknowledgement}
\section{Data availability statement}
The data that supports the findings of this study is available from the corresponding author upon reasonable request.
\section{Code availability statement}
The code to generate the transient-absorption spectra is available under \url{https://github.com/psebastianzjl/pushpulse_single_trajectory}.
\section{Competing interests}
The authors declare that they have no competing interests.
%
%
\bibliography{main}
\end{document}